\documentclass[a4paper]{article}

\def\const{\mbox{const}}

\def\l{\left(}
\def\r{\right)}
\def\la{\langle }
\def\ra{\rangle }

\usepackage{amsfonts, amsmath, amssymb}

\newcommand{\be}{\begin{equation}}
\newcommand{\ee}{\end{equation}}
\newcommand{\bea}{\begin{eqnarray}}
\newcommand{\eea}{\end{eqnarray}}
\newcommand{\bg}{\begin{gather}}
\newcommand{\eg}{\end{gather}}
\newcommand{\bseq}{\begin{subequations}}
\newcommand{\eseq}{\end{subequations}}

\newcommand{\Tr}{{\rm Tr}}
\def\half{\frac{1}{2}}
\newcommand{\eq}[1]{(\ref{#1})}

\newcommand{\bra}[1]{\langle #1 |}
\newcommand{\ket}[1]{| #1 \rangle}

\newcommand{\upa}{ {\uparrow}   }
\newcommand{\doa}{ {\downarrow} }

\begin{document}

\author{
F.L.~Levkovich-Maslyuk$^{a}$\footnote{flev@ms2.inr.ac.ru}\\
$^a${\small  Moscow State University, 119991, Moscow, Russia}\\
}

\title{Two destructive effects of decoherence on Bell inequality violation.}
\date{}
\maketitle

\begin{abstract}
We consider a system of two spin-$\frac{1}{2}$ particles, initially in an entangled Bell state. If one of the particles is interacting with an environment (e.g., a collection of $N$ independent spins), the two-particle system undergoes decoherence. Using a simple model of decoherence, we show that this process has two consequences. First, the maximal amount by which the CHSH inequality is violated decays to zero. Second, the set of directions of measurement for which the inequality is violated is reduced in the course of decoherence. The volume of that set is bounded above by $\mathrm{const}\cdot |r|^2$, where $r$ is the decoherence factor. We obtain similar results for the case when each of the two particles is in interaction with a separate environment.
\end{abstract}

In a model of local hidden variables (LHVs) the statistical correlations of measurements on a composite physical system must obey certain bounds, called Bell inequalities~\cite{origBell, origCHSH}. It is well known that for some quantum systems the observables can be chosen in such a way that at least one of the inequalities is violated~\cite{origBell, origCHSH, Cirelson}. Therefore, certain quantum systems cannot be described by an LHV model. Clauser, Horne, Shimony and Holt (CHSH)~\cite{origCHSH} obtained a Bell type inequality which provided a way to experimentally test the existence of nonlocal correlations for such systems. For a pair of spin-$\half$ particles (such a particle can represent a qubit) the CHSH inequality can be written in the following form:
\be
\label{CHSHineqWithE}
| E(\mathbf a, \mathbf b) + E(\mathbf a', \mathbf b)
+ E(\mathbf a, \mathbf b') - E(\mathbf a', \mathbf b')| \le 2.
\ee
Here, $E(\mathbf a, \mathbf b)$ denotes the expectation value of the product
$A_1(\mathbf a) \cdot A_2(\mathbf b)$, where $A_k(\mathbf a)$ is the result of a measurement of the $k$th particle's spin projection in the direction $\mathbf a$ ($A_k$ takes the values of
$\pm 1$). Inequality~\eq{CHSHineqWithE} must hold if the results of spin measurements are described by an LHV model. Moreover, if Eq.~\eq{CHSHineqWithE} is not violated for arbitrary vectors
$\mathbf a, \mathbf a', \mathbf b, \mathbf b'$,
then there exists~\cite{Fine82Criterion} an LHV model describing the results which are obtained when a single ideal measurement is performed on each of the particles.

Consider now the singlet state of the two-particle system:
\be
	\label{bellState}
	\ket{\psi} = \frac{1}{\sqrt{2}}
	\l \ket{\upa}\ket{\doa} - \ket{\doa}\ket{\upa} \r,
\ee
where $\ket{\upa}$ and $\ket{\doa}$ denote, respectively, the states with spin up or down along the $Z$ axis. The operator which corresponds to the left-hand side of the CHSH inequality is given by
\be
	\label{BCHSH}
	B_{\mathrm{CHSH}} = \mathbf a  \cdot \sigma \otimes \l\mathbf {b+b'}\r \cdot \sigma +
							  \mathbf a' \cdot \sigma \otimes \l\mathbf {b-b'}\r \cdot \sigma,
\ee
where $\mathbf a, \mathbf a', \mathbf b, \mathbf b'$ are unit vectors in $\mathbb R^3$, $\sigma _i$ are the Pauli matrices, $\mathbf a  \cdot \sigma = \sum\nolimits_{i=1}^3 a_i \sigma _i$. Hence, the CHSH inequality  can be written as
$|\la\psi|B_{\mathrm{CHSH}}|\psi\ra| \le 2$.
State~\eq{bellState} does not admit an LHV model, as the vectors
$\mathbf a, \mathbf a', \mathbf b, \mathbf b'$
can be chosen in such a way that
\be
	|\la\psi|B_{\mathrm{CHSH}}|\psi\ra| = 2\sqrt{2}.
\ee
Thus, the CHSH inequality is violated by the maximal amount possible for any state~\cite{Cirelson}.  The volume of the set of all inequality-violating vectors
$\mathbf a, \mathbf a', \mathbf b, \mathbf b'$ is significantly nonzero for
state~\eq{bellState} (we will show this below).

Now suppose that one of the two particles is in interaction with a many-particle environment. This interaction can cause the two-particle system to undergo decoherence~\cite{Zurek82Einselection, ZurekPhysToday91, ZurekRevModPhys03}, and the system will practically lose its quantum properties. The decay of Bell violation and entanglement measures of a decohering quantum system was studied recently in a number of works (see, e.g.,~\cite{Eberly03, Miranowicz04, Dodd04, Hein05, Li05}).
In this paper we show that in the model of decoherence we use, there are two ways in which partial decoherence affects the ability of our two-particle system to violate the CHSH inequality.
First, the maximal amount by which the CHSH inequality can be violated for the system becomes of order $|r|^2$ for small $|r|$, where $r$ is the decoherence factor.
In addition, the volume of the set of vectors
$\mathbf a, \mathbf a', \mathbf b, \mathbf b'$
for which the inequality is violated is bounded above by $\const \cdot |r|^2$ and hence tends to zero as $r \to 0$.

We also consider the case when each of the two particles interacts with its own independent environment. Decoherence of the $k$th particle is then characterized by the decoherence factor $r_k$ ($k = 1,2$). In this case, we obtain the same estimates as above for the maximal violation and the volume of the set of CHSH inequality-violating vectors, with $|r|$ replaced by $|r_1 r_2|$.

\bigskip
We assume that the interaction which causes decoherence is such that $\ket{\upa}$ and $\ket{\doa}$ are the pointer states~\cite{ZurekRevModPhys03} for the particle $\mathcal P$ which is interacting with the environment $\mathcal E$. This is the case, for example, in the spin-spin model which was studied by Zurek and co-workers~\cite{Zurek82Einselection, Zurek05SpinSpin}, if the self-Hamiltonians of $\mathcal P$ and $\mathcal E$ are neglected. The evolution of the system can then be described as follows. Suppose the combined $\mathcal P + \mathcal E$ initial state has the form
\be
	\ket{\Psi_{\mathcal P \mathcal E} (0)} = (a\ket{\upa} + b\ket{\doa})  \ket{\mathcal E _{init}}.
\ee
Then the state at an arbitrary time is given by
\be
	\ket{\Psi_{\mathcal P \mathcal E} (t)} = a\ket{\upa}\ket{\mathcal E _{\upa}(t)} +
														  b\ket{\doa}\ket{\mathcal E _{\doa}(t)},
\ee
where the decoherence factor 
\be
\label{decohFacDef}
	r(t) = \la\mathcal E _{\doa}(t)|\mathcal E _{\upa}(t)\ra 
\ee
decays to zero as $t$ increases. The reduced density matrix of $\mathcal P$ has the form
\be
	\begin{split}
	\rho_{\mathcal P} &=\Tr_{\mathcal E}
								\ket{\Psi_{\mathcal P \mathcal E} (t)}
								\bra{\Psi_{\mathcal P \mathcal E} (t)} \\
							&= |a|^2\ket{\upa}\bra{\upa} + ab^*r(t)\ket{\upa}\bra{\doa} \\
							&+ a^*br^*(t)\ket{\doa}\bra{\upa} + |b|^2\ket{\doa}\bra{\doa}.
	\end{split}
\ee
Therefore, for ${ |r(t)| \ll 1 }$, the matrix $\rho_{\mathcal P}$ is approximately diagonal in the pointer-state basis ${ \{ \ket{\upa}, \ket{\doa} \} }$. Thus, in the process of decoherence the state of $\mathcal P$ becomes almost indistinguishable from a classical mixture of the pointer states $\ket{\upa}, \ket{\doa}$ (with respective probabilities $|a|^2$ and $|b|^2$).

\bigskip
We now proceed to the derivation of our main results. Let $\mathcal S$ denote a system of two spin-$\half$ particles, initially in Bell state~\eq{bellState}. First, suppose the second particle is interacting with an environment $\mathcal E$ in the way just described, which causes the system to undergo decoherence. Then the combined $\mathcal S + \mathcal E$ state at time $t$ is given by
\be
	\ket{\Psi_{\mathcal S \mathcal E} (t)} =
	\frac{1}{\sqrt{2}} \bigl(
	\ket{\upa}\ket{\doa}\ket{\mathcal E _{\doa}(t)} -
	\ket{\doa}\ket{\upa}\ket{\mathcal E _{\upa}(t)} \bigr).
\ee
Hence, the reduced density matrix of $\mathcal S$ in the basis~
$\bigl\{\ket{\upa}\ket{\upa}, \ket{\upa}\ket{\doa},
\ket{\doa}\ket{\upa}, \ket{\doa}\ket{\doa}\bigr\}$
has the form
\be
	\label{rhoTwoSpinsExact}
	\rho = \half
	\begin{pmatrix}
	0 & 0 & 0 & 0 \\
	0 & 1 & -r^* & 0 \\
	0 & -r & 1 & 0 \\
	0 & 0 & 0 & 0
	\end{pmatrix}
	,
\ee
where $r$ is defined by Eq.~\eq{decohFacDef}.

Now consider the case when each particle of the pair is in interaction with its own environment. Denote by $\mathcal E$ and $\mathcal E'$ the environments of the first and the second particle, respectively. We assume that $\mathcal E$ and $\mathcal E'$ do not interact with each other. The state of
$\mathcal S + \mathcal E + \mathcal E'$ at time $t$ can then be written as
\be
\label{stateTwoEnvirons}
	\ket{\Psi_{\mathcal S \mathcal E \mathcal E'} (t)} =
	\frac{1}{\sqrt{2}} \Bigl[
	\ket{\upa} \ket{\doa} \ket{\mathcal E _{\upa} (t)} \ket{\mathcal E _{\doa} ' (t)} -
	\ket{\doa} \ket{\upa} \ket{\mathcal E _{\doa} (t)} \ket{\mathcal E _{\upa} ' (t)}
	\Bigr].
\ee
The decoherence factors for the first and second particles are given by
\be
\label{twoDecohFactors}
	r_1(t) = \la\mathcal E _{\doa}   (t)|\mathcal E _{\upa}   (t)\ra, \ 
	r_2(t) = \la\mathcal E _{\doa} ' (t)|\mathcal E _{\upa} ' (t)\ra.	
\ee
For state~\eq{stateTwoEnvirons}, the reduced density matrix of $\mathcal S$ has form~\eq{rhoTwoSpinsExact}, with $r$ replaced by $r_1^* r_2$. We see that the two-particle system with decoherence due to interactions of both particles is described by the same density matrix as the system in which only one particle interacts with its environment.

For a pair of photons with a similar density matrix, maximal violation of the CHSH inequality was computed in~\cite{Cabello05}, although an analytic expression for the maximal violation was not given in that work. 
For our system the CHSH inequality can be written as
\be
	\label{CHSHineqB}
	|\la B_{\mathrm{CHSH}} \ra _{\rho}| \le 2,
\ee
where
$\la B_{\mathrm{CHSH}} \ra _{\rho} = \Tr \l\rho B_{\mathrm{CHSH}}\r$, with $\rho$ given by Eq.~\eq{rhoTwoSpinsExact} and the operator $B_{\mathrm{CHSH}}$ given by Eq.~\eq{BCHSH}.
Following~\cite{hor3CriterionCHSH}, we represent $\rho$ in the form
\be
	\rho = \frac{1}{4}
	\l
	I\otimes I + \mathbf r \cdot \sigma\otimes I + I\otimes \sigma \cdot \mathbf s +
	\sum _{n,m=1}^3 t_{nm}\sigma_i\otimes\sigma_i
	\r,
\ee
where $I$ is the identity matrix. Let $T_{\rho}$ be the matrix formed by the coefficients $t_{nm}$. Then
\be
\label{BwithTrho}
	\la B_{\mathrm{CHSH}} \ra_{\rho} =
	\l \mathbf a ,T_{\rho}(\mathbf {b+b'}) \r +
	\l \mathbf a',T_{\rho}(\mathbf {b-b'}) \r.
\ee
As $t_{nm}=\Tr\l\rho\sigma_n\otimes\sigma_m\r$, we find that for density
matrix~\eq{rhoTwoSpinsExact},
\be
\label{Trho}
	T_{\rho} =
	\begin{pmatrix}
	-\mathrm{Re}(r) & \mathrm{Im}(r) & 0 \\
	-\mathrm{Im}(r) & -\mathrm{Re}(r) & 0 \\
	0 & 0 & -1
	\end{pmatrix}.
\ee
Introduce the matrix
$U_{\rho} = T_{\rho}^{\mathrm T} T_{\rho}$,
where $T_{\rho}^{\mathrm T}$ is the transposition of $T_{\rho}$, and let $M(\rho)$ be the sum of the two largest eigenvalues of $U_{\rho}$. It was shown in~\cite{hor3CriterionCHSH} that for a given density matrix the maximal value of $|\la B_{\mathrm{CHSH}} \ra_{\rho}|$ with respect to the vectors
$\mathbf a, \mathbf a', \mathbf b, \mathbf b'$
is
$2\sqrt{M(\rho)}$.
For density matrix~\eq{rhoTwoSpinsExact} we have
\be
	M(\rho) = 1 + |r|^2.
\ee
Therefore, for our system of two spin-$\half$ particles we get
\be
\label{maxViolationWithr}
	\max\limits_{\mathbf a, \mathbf a', \mathbf b, \mathbf b'}
	|\la B_{\mathrm{CHSH}} \ra_{\rho}| = 2\sqrt{1+|r|^2}.
\ee
Thus, the maximal value by which CHSH inequality~\eq{CHSHineqB} can be violated for our system is of order $|r|^2$ for small $|r|$. For $r=0$ the inequality is not violated regardless of the choice of vectors $\mathbf a, \mathbf a', \mathbf b, \mathbf b'$. Replacing $r$ by $r_1^* r_2$, with $r_{1,2}$ defined by Eq.~\eq{twoDecohFactors}, we find that in the case of decoherence due to interactions of both particles, the maximal violation is of order $|r_1 r_2|^2$ (for small $|r_1 r_2|$).

\bigskip
We will now show that in the process of decoherence the set of directions of measurement for which the CHSH inequality is violated becomes reduced in comparison with that set for the initial state. First let us introduce some notation. For our density matrix~\eq{rhoTwoSpinsExact},
$\la B_{\mathrm{CHSH}} \ra _{\rho}$
is a function of
$\mathbf a, \mathbf a', \mathbf b, \mathbf b'$
and $r$. Its domain is $V \times \mathbb C$, where $V$ is the product of four unit spheres:
$V = S^2 \times S^2 \times S^2 \times S^2$.
Denote by $L(r)$ the set of CHSH inequality-violating vectors:
\be
	L(r) =
	\bigl\{
	\l\mathbf a, \mathbf a', \mathbf b, \mathbf b'\r \in V
	\ \bigl| \bigr. \ 
	|\la B_{\mathrm{CHSH}} \ra _{\rho}| > 2
	\bigr\}.
\ee
Denote by $Vol$ the natural measure on $V$ obtained from the measure which describes area on $S^2$.
It will be shown below that
\be
	\label{upperBoundOnL}
	Vol \left[L(r) \right] \le \const \cdot |r|^2.
\ee
Hence, $Vol \left[L(r) \right] \to 0$ as $r \to 0$. For the case when each particle of the pair is in interaction with its environment, inequality~\eq{upperBoundOnL} takes the form
\be
	Vol \left[L(r_1, r_2) \right] \le \const \cdot |r_1 r_2|^2.
\ee

On the other hand, it is easy to see that for initial state~\eq{bellState} the measure of the set of CHSH inequality-violating vectors is significantly nonzero. To show this, we first construct a continuous family of vectors for which the inequality is maximally violated, i.e.,
$|\la B_{\mathrm{CHSH}} \ra| = 2\sqrt{2}$. This can be done as follows. Consider any two mutually orthogonal unit vectors $\mathbf a \perp \mathbf a'$, and take
\be
	\mathbf b  = \frac{\mathbf a + \mathbf a'}{\sqrt{2}}, \ \ 
	\mathbf b' = \frac{\mathbf a - \mathbf a'}{\sqrt{2}}.
\ee
As state~\eq{bellState} is described by density matrix~\eq{rhoTwoSpinsExact} with $r=1$,
from Eq.~\eq{Trho} we have
	$T_{\rho} = -I$,
where $I$ is the identity matrix. Using Eq.~\eq{BwithTrho}, we find that
	$\la B_{\mathrm{CHSH}} \ra = 2\sqrt{2}$
for the chosen vectors
	$\mathbf a, \mathbf a', \mathbf b, \mathbf b'$,
which form a continuous family. From Eq.~\eq{BwithTrho} we see that
	$\la B_{\mathrm{CHSH}} \ra$
is a quadratic form of the components of unit vectors
	$\mathbf a, \mathbf a', \mathbf b, \mathbf b'$.
Hence, the gradient of $\la B_{\mathrm{CHSH}} \ra$ is bounded above on $V$. Therefore, the family of vectors we just constructed has a neighbourhood of nonzero volume in which
$\la B_{\mathrm{CHSH}} \ra > 2$.
Thus the set of inequality-violating vectors for state~\eq{bellState} is of significantly nonzero measure. From Eq.~\eq{upperBoundOnL} we conclude that in the process of decoherence the set
$L(r)$ is greatly reduced.

We will now prove inequality~\eq{upperBoundOnL}. It can be found from Eqs.~\eq{BwithTrho} and ~\eq{Trho} that
\be
\label{BwithZP}
	\la B_{\mathrm{CHSH}} \ra_{\rho} = Z + |r| P,
\ee
where the quantities $Z$ and $P$ are given by
\be
	Z = -a_z(b_z+b'_z) - a'_z(b_z-b'_z),
\ee
\be
\label{defP}
	P = 
	\left[
	\mathbf  a_{\parallel} \cdot \hat \alpha \l\mathbf b_{\parallel}+\mathbf b'_{\parallel}\r + 
	\mathbf a'_{\parallel} \cdot \hat \alpha \l\mathbf b_{\parallel}-\mathbf b'_{\parallel}\r
	\right].
\ee
Here, $\mathbf a_{\parallel}$ denotes the projection of the vector $\mathbf a$ onto the $xy$ plane, and
$\hat \alpha$ is the operator of rotation by angle $\alpha$ in that plane, with the angle defined by
\be
	\cos \alpha = -\frac{\mathrm{Re}(r)}{|r|}, \;\;
	\sin \alpha = -\frac{\mathrm{Im}(r)}{|r|}.
\ee
From Eq.~\eq{defP} we have $|P| \le 2\sqrt{2}$. Hence, from Eq.~\eq{BwithZP} we obtain
\be
	|\la B_{\mathrm{CHSH}} \ra _{\rho}| \le |Z| + 2\sqrt{2}|r|.
\ee
Therefore, if $\l\mathbf a, \mathbf a', \mathbf b, \mathbf b'\r \in L(r)$, then
$|Z| > 2 - 2\sqrt{2}|r|$. Thus, introducing the set
\be
	E(r) =
	\bigl\{
	\l\mathbf a, \mathbf a', \mathbf b, \mathbf b'\r \in V
	\ \bigl| \bigr. \ 
	|Z| > 2 - 2\sqrt{2}|r|
	\bigr\},
\ee
we have $L(r) \subset E(r)$ and
\be
\label{VolLleVolE}
	Vol \left[L(r) \right] \le Vol \left[E(r) \right].
\ee
Note that the condition $|Z| > 2 - 2\sqrt{2}|r|$ which defines $E(r)$ does not include $P$. This allows
us to obtain an upper bound on the measure of $E(r)$ in the following way.
One can show that if $0 < k < 1, \ |a_z| \le k$, and $|a'_z| \le k$, then $|Z| \le 2k$. Choosing
$k = 1 - \sqrt{2}|r|$, we find\footnote
	{In the proof of inequality~\eq{upperBoundOnL} we may assume that
	$\sqrt{2}|r| < 1$ and hence $k > 0$.}
that if $\l\mathbf a, \mathbf a', \mathbf b, \mathbf b'\r \in E(r)$, then
\be
	\label{aIneqForEr}
	\begin{aligned}
	|a_z|  > 1 - \sqrt{2}|r| & \ \ \ \mathrm{or} \\
	|a'_z| > 1 - \sqrt{2}|r| &.
	\end{aligned}
\ee
Similarly, if $\l\mathbf a, \mathbf a', \mathbf b, \mathbf b'\r \in E(r)$,
then also
\be
	\label{bIneqForEr}
	\begin{aligned}
	|b_z|  > 1 - \sqrt{2}|r| & \ \ \ \mathrm{or} \\
	|b'_z| > 1 - \sqrt{2}|r| &.
	\end{aligned}
\ee
Denote by $S_{\delta}$ that part of the unit sphere $x^2 + y^2 + z^2 = 1$ where the condition
$|z| > 1 - \delta$
holds.
From Eqs.~\eq{aIneqForEr} and~\eq{bIneqForEr}, we see that $E(r)$ can be included into a union of four sets, each of which is the product of two entire spheres $S^2$ and two truncated spheres $S_{\delta}$
(with $\delta = \sqrt{2}|r|$).
Those four sets have equal measures
$A(\sqrt{2}|r|) \cdot A(\sqrt{2}|r|) \cdot 4\pi \cdot 4\pi$, where $A\l\delta\r$ is the area of the set $S_{\delta}$ on the unit sphere. Therefore,
\be
	Vol \left[E(r) \right] \le
	\const \cdot
	\l \, A \l\sqrt{2}|r|\r \, \r ^2.
\ee
Hence, as $A\l\delta\r$ is linear in $\delta$, from Eq.~\eq{VolLleVolE} we obtain inequality~\eq{upperBoundOnL}.

\bigskip

In conclusion, we have shown that in the case of decoherence in one particle of the Bell pair, the maximal value by which the CHSH inequality can be violated is of order $|r|^2$ for small $|r|$, where $r$ is the decoherence factor. Moreover, the volume of the set of inequality-violating directions of measurement is bounded above by $\const \cdot |r|^2$ and tends to zero as $r \to 0$. The estimates obtained apply also to a system with independent decoherence due to interactions of both particles, with $r$ replaced by the product of the two decoherence factors. These results show that as decoherence progresses, the nonlocality expressed in the violation of the CHSH inequality becomes weaker in two complementary ways. An interesting question is whether such conclusions remain true for other models of decoherence, e.g., amplitude damping or depolarization. 

I am grateful to V. A. Rubakov for setting this problem and for useful discussions.

\end{document}